# High-precision astrometric studies in direct imaging with SPHERE

Anne-Lise Maire[1,2]
Gaël Chauvin[3]
Arthur Vigan[4]
Raffaele Gratton[5]
Maud Langlois[6]
Julien H. Girard[7]
Matthew A. Kenworthy[8]
Jörg-Uwe Pott[2]
Thomas Henning[2]
Pierre Kervella[9]
Sylvestre Lacour[9]
Emily L. Rickman[7]
Anthony Boccaletti[9]
Philippe Delorme[3]
Michael R. Meyer[10]
Mathias Nowak[11]
Sascha P. Quanz[12]
Alice Zurlo[13]

[1] University of Liège, Liège, Belgium
[2] Max-Planck-Institut für Astronomie, Heidelberg, Germany
[3] Institut de Planétologie et d'Astrophysique de Grenoble, France
[4] Aix Marseille Univ, CNRS, CNES, LAM, Marseille, France
[5] INAF-Osservatorio Astronomico di Padova, Padova, Italy
[6] Centre de Recherche Astrophysique de Lyon, Saint Genis Laval, France
[7] Space Telescope Science Institute, Baltimore, USA
[8] Leiden Observatory, Leiden, Netherlands
[9] Observatoire de Paris-Meudon, Meudon, France
[10] University of Michigan, Ann Arbor, USA
[11] University of Cambridge, Cambridge, UK
[12] Eidgenössische Technische Hochschule Zürich, Switzerland
[13] Universidad Diego Portales, Santiago, Chile

**Orbital monitoring of exoplanetary and stellar systems is fundamental for analysing their architecture, dynamical stability and evolution, and mechanisms of formation. Current high-contrast extreme-adaptive optics imagers like SPHERE, GPI, and SCExAO+CHARIS explore the population of giant exoplanets and brown dwarf and stellar companions beyond typically 10 au, covering generally a small fraction of the orbit (<20%) leading to degeneracies and biases in the orbital parameters. Precise and robust measurements over time of the position of the companions are critical, which require good knowledge of the instrumental limitations and dedicated observing strategies. The homogeneous dedicated calibration strategy for astrometry implemented for SPHERE has facilitated high-precision studies by its users since its start of operation in 2014. As the precision of exoplanet imaging instruments is now reaching milliarcseconds and is expected to improve**

**with the upcoming facilities, we initiated a community effort, triggered by the SPHERE experience, to share lessons learned for high-precision astrometry in direct imaging. A homogeneous strategy would strongly benefit the VLT community, in synergy with VLTI instruments like GRAVITY/GRAVITY+, future instruments like ERIS and MAVIS, and in preparation for the exploitation of the ELT's first instruments MICADO, HARMONI, and METIS.**

## Motivation

High-precision relative astrometry in direct imaging is critical for various and unique science cases beyond determining the orbital parameters of exoplanets, brown dwarf companions or multiple stellar systems. For exoplanet surveys (Langlois et al., 2021), it is instrumental for testing the nature of the faint sources detected near the targeted stars (Figure 1, left). The fields of view used are typically too small for absolute astrometry, so that astrometry relative to the targeted star is used. Multiple-epoch monitoring enables to test if the candidate companions are comoving with similar proper and parallactic motion than the host star by rejecting contamination by stationary (or slowly moving with the local field) background or foreground source. More precise measurements allow for faster confirmations, which is critical given the international competition. This approach requires a second observation, e.g., from archival data. A word of caution is the possibility of a candidate companion having significant proper motion and mimicking a physical companion with orbital motion (e.g., HD131399b; Nielsen et al., 2017). Multiple-epoch monitoring remains the most reliable approach to confirm a candidate companion, and ultimately to resolve its orbital motion to confirm that it is gravitationally bound.

Constraining the orbital parameters of a companion provides clues on its formation and dynamical history. Orbits with small eccentricities are consistent with planetary formation within circumstellar discs, and similar to the Solar System's configuration. Larger eccentricities might be connected to stellar-like formation mechanisms or indicative of subsequent dynamical planet - planet interactions in multiple planetary systems that could explain the broad eccentricity distribution of exoplanets detected with the radial velocity technique. Another valuable output of orbital fits are predictions for the positions. This is important to optimise follow-up observations at longer wavelengths (lower angular resolution) or with slit/fibre spectrometry.

There is a strong synergy between direct imaging and radial velocities and absolute astrometry for orbital fits. Firstly, it can constrain the mass of the companions, which is a fundamental step towards the calibration of models of the evolution of young giant planets, brown dwarfs, and low-mass stars. For imaged companions, most mass measurements come from evolutionary models, which suffer from large theoretical uncertainties (e.g., clouds and molecular opacities for the atmosphere, initial entropy for the formation). Secondly, it allows for the breaking of degeneracies in the orbital parameters. Radial velocities are degenerate with the inclination (essential to constrain the mass), but are lifted with imaging and absolute astrometry. For multiple-companion systems, direct imaging is valuable to break the degeneracies with radial velocities or absolute astrometry due to the unknown orbital phases, although the analysis of the dynamical stability may also be used. Thanks to the 24-yr baseline between Hipparcos and Gaia DR2, absolute astrometry can now detect massive substellar companions at the separations probed by direct imaging. The bridging of these techniques will increase with Gaia and the ELT to closer-in and/or planetary-mass companions, with the prospect of a complete view of planetary and stellar systems.

Direct imaging offers a unique means to simultaneously analyse companions and their birth environment, the circumstellar discs. Determining the orbit of the companions gives insights on potential dynamical interactions. Such systems provide valuable benchmarks for planet formation and migration models. The analysis of companion-disc dynamical interactions will also help to clarify which disc features (e.g., spiral arms, rings, clumps) can be reliably associated with companions. Another research field that has recently emerged is the monitoring of the motion of disc features to discriminate between different production mechanisms (Figure 1, right). For instance, misaligned inner discs or close-in companions have been proposed to explain shadows cast on the outer discs in various protoplanetary discs.

## The problem of astrometric biases

The advent of the first dedicated exoplanet imaging instruments (SPHERE, GPI, SCExAO+CHARIS; e.g., Beuzit et al. 2019) has improved the precision of relative astrometric measurements of young substellar companions, from about 10 mas to about 1-2 mas.

Measurements with higher precision are more sensitive to underestimated biases - these can be caused by different methods for the data analysis and/or calibration, our limited knowledge of the thermo-mechanical stability of the instruments, and the use of different instruments (e.g., after upgrades). Given the long orbital periods of the imaged companions compared to the lifetime of instruments, maximising the measured orbital arc is vital to derive more robust orbital constraints. Underestimated biases may also affect comotion tests of candidate companions and trigger follow-up observations by mistake, wasting telescope time.

Figure 2 illustrates the importance of a good knowledge of the biases in comotion tests of candidate companions using different instruments. For a star with many candidate companions, the biases can be estimated assuming that most of them are background contaminants. For a star with a single candidate companion, a new observation is required to conclude.

Figure 3 illustrates the importance of a good knowledge of the biases in orbital fits for the exoplanet HIP65426b (Chauvin et al., 2017; Cheetham et al., 2019). Low eccentricities and a bimodal distribution for the time at periapsis are favoured when combining SPHERE and NaCo data obtained in 2016-2017, whereas the eccentricity is not well constrained and can be high and the periapsis is in the future when fitting SPHERE data obtained in 2016-2018. These discrepant results point to underestimated systematic uncertainties between the SPHERE and NaCo data.

## SPHERE astrometric strategy

A homogeneous and regular astrometric calibration is critical to minimise the biases and analyse the astrometric stability over time. A good astrometric stability eases comotion tests and orbital monitoring of imaged companions. It relaxes the need to take calibration data close to the science observations and reduces the calibration overhead at the telescope.

The astrometric strategy for the SPHERE SHINE survey was devised by the consortium before the commissioning and subsequently refined. It relies on: 1/ an observing procedure for a precise determination of the star location behind the coronagraph (Langlois et al., 2021), 2/ an accurate determination of the instrument overheads and metrology, and 3/ regular observations of fields in stellar clusters for the astrometric calibration (Figure 4; Maire et al., 2016).

We chose fields in stellar clusters as main astrometric calibrators because the large number of stars available allows for precise measurements. They also allow for measuring the distortion from the telescope optics. We selected cluster fields with positions measured precisely by the Hubble Space Telescope, which has a good absolute calibration. We further selected fields with a bright star for adaptive optics (AO) guiding (R<~13.5 mag). Finally, we repeatedly observed two fields to cover the whole year, 47 Tucanae and NGC3603. We chose 47 Tucanae as the reference field because the catalog provides the stellar proper motions (Bellini et al., 2014). Langlois et al. (2021) compared the relative astrometry for widely-separated and bright candidate companions observed with SPHERE and present in Gaia DR2 catalogue. The mean offset in separation is -2.8±1.5 mas (3.9 mas rms) and in position angle 0.06±0.04 deg (0.11 deg rms). The rms agree well with the expected uncertainties in these quantities in SPHERE data.

To analyse the astrometric data and derive the calibration, we developed a tool (Maire et al., 2016), included in the SPHERE Data Centre[1]. The distortion is mainly due to the optics in SPHERE and is stable in time (see Table). It produces differences in the horizontal and vertical pixel scales which amount to 6 mas at 1 arcsec. The astrometric requirements are 5 mas (goal 1 mas).

Figure 5 shows the temporal evolution of the pixel scale and correction angle to the North (Maire et al., in preparation). Except for pixel scale measurements obtained during the commissioning, SPHERE has demonstrated a remarkable astrometric stability over five years. The standard deviation for the pixel scale measured on 47 Tucanae is 0.004 mas/pixel and for the correction angle to the North 0.04 deg. These variations translate into uncertainties at 1 arcsec of 0.33 and 0.70 mas, respectively, within the baseline astrometric requirements. We plan to release the measurements in the SPHERE Target Data Base[2].

The pixel scale and correction angle to the North have also been monitored in the ESO monthly calibration plan. The SHINE astrometric fields have been observed without coronagraph. We analysed the data at the SPHERE Data Centre to compute an astrometric table for the reduction of the open-time data. About 80% of the observations were not suitable for deriving a good calibration. Work is ongoing with the ESO staff to improve the setup of their observations.

The astrometric calibration of the SPHERE images also requires measurement of the angle offset of the pupil in pupil-tracking mode. The pupil-tracking mode allows for subtracting the aberrations in the images due to the telescope and instrument. We monitored this parameter in 2014-2016 and showed that it is stable. Work is ongoing to monitor it in the ESO calibration plan.

In contrast, the astrometric calibration for NaCo was heterogeneous, irregular, and mostly left to the observing teams. NaCo also underwent technical interventions to commission new observing modes or fix issues, and was moved to another UT. This resulted in poor astrometric stability, making the use of the data for high-precision relative astrometry more difficult. The limitations encountered with NaCo were taken into account in the astrometric strategy of SPHERE.

| | |
|---|---|
| Requirements separation (mas) | 5 (goal: 1) |
| Requirements position angle (deg) | 0.2 |
| | |
| Achieved precision calibration pixel scale at 1'' (mas) | 0.33 |
| Achieved precision calibration distortion at 1'' (mas) | 0.2 |
| Achieved precision calibration angle to North (deg) | 0.04 |
| Achieved precision calibration angle pupil tracking (deg) | 0.06 |

## Highlights from SPHERE results

Thanks to the good astrometric precision and stability of SPHERE, most users rely on the calibration derived by the instrument consortium. SPHERE has allowed for the discovery of 15 substellar and stellar companions next to stars. It has been used for about 20 orbital studies, in combination with other imaging, radial velocity, and/or absolute astrometric measurements.

The orbital analyses of the exoplanets β Pictoris b (Lagrange et al., 2019) and 51 Eridani b (Maire et al., 2019) are good examples where biases between different instruments had to be dealt with. β Pictoris b was monitored with NaCo then SPHERE. It was recovered in September 2018 after conjunction with the star. The SPHERE data are now probing the North-East part of the orbit, which was only covered by one NaCo measurement in 2003, and favour low eccentricities. 51 Eridani b was monitored for three years. Coupled with GPI data, orbital curvature was detected in this system for the first time and the fit suggests a high eccentricity (~0.3-0.6). A high eccentricity hints at dynamical interactions that perturbed the orbit of the planet, possibly by another as yet undetected planet.

The orbital predictions were also used for GRAVITY observations to get spectra at longer wavelengths and higher resolutions (2.0-2.4 µm, R~500) compared to SPHERE (1.0-2.3 µm, R~50) and to get exquisite astrometry (~30 times more precise), confirming the robustness of the SPHERE calibration plan. Companion-disc dynamical interactions were studied for several systems, including systems with brown dwarfs within the cavity of debris discs. HR2562B could carve the disc cavity, whereas another companion may be needed around HD206893. Disc features were monitored, such as the arch-like features moving away from the star AU Microscopii. The current scenario involves dust produced by an unseen parent body and expelled by the stellar wind. The rotation of the spiral arms of MWC758 was shown to be compatible with a planet-driven mechanism.

Future astrometric studies with direct imaging facilities at ESO

Pursuing the monitoring of known companions and disc features will be important to refine their orbit and their formation mechanisms, respectively. Moreover, Gaia is expected to detect a large number of giant exoplanets. Young exoplanets detected from acceleration measurements will be prime targets for imaging to confirm and firmly constrain their orbit and mass. This large sample of exoplanets beyond a few au will allow for statistical analyses of the distributions of eccentricities and relative inclinations to the stellar equatorial planes (for multiple-planet systems also mutual inclinations). Such analyses will be crucial to understand their formation and evolution, and the relation between planet and binary star formation mechanisms.

The next step for exoplanet imaging will be the Extremely Large Telescope and its first three instruments: MICADO, HARMONI, and METIS. They will access smaller separations to stars down to 1 au to detect predominantly giant exoplanets. Due to the combination of increased angular resolution and collecting aperture, diffraction-limited ELT observations will at the same time access smaller angular separations, and achieve higher astrometric precision at angular separations accessible to 8m-class imagers. MICADO and HARMONI will be sensitive to young planets, whereas METIS will reach mature planets. Before the ELT, ERIS, GRAVITY+, and a potential SPHERE upgrade will be operational on the VLT/I. ERIS will be suitable for imaging giant exoplanets around young stars, more mature giant exoplanets which are too faint for the SPHERE AO system. GRAVITY+ will reach better sensitivity than GRAVITY to access mature exoplanets.

A joint and homogeneous strategy shared by the exoplanet imaging facilities at ESO will enhance their use for high-precision astrometry, by minimizing biases. The successful calibration plan implemented for SPHERE could be applied and adapted to these instruments. If proposed by the consortia of future instruments, interactions with ESO would be valuable to check if such a calibration plan could be adopted. As SPHERE is expected to be operational until the first years of ELT operations, parallel observations could be used to check the astrometric consistency. GRAVITY could be used to test/validate the absolute calibration of coronagraphic instruments, thanks to the absolute calibration provided by its internal metrology system (Lacour et al., 2014).

We recently started an initiative between the SPHERE team and the teams in charge of the high-contrast imaging modes of upcoming ESO exoplanet imaging facilities at VLT/I and ELT to share the SPHERE experience and lessons learned in the field of astrometric characterization of exoplanets and discs. We firmly believe that it offers the opportunity to federate our community: 1/ to revisit past studies through archival data mining, 2/ to push the calibration strategy and performance of current instruments in operation, and 3/ to share this expertise with consortia of forthcoming instruments at VLT/I and ELT to optimally prepare their scientific exploitation. We envision the preparation of a workshop on this topic in the future.

Acknowledgements
A.L.M. acknowledges financial support from the European Research Council under the European Union's Horizon 2020 research and innovation program (Grant Agreement No. 819155). We thank Leonard Burtscher for useful comments. This work has made use of the SPHERE Data Centre, operated by OSUG/IPAG (Grenoble), PYTHEAS/LAM/CeSAM (Marseille), OCA/Lagrange (Nice), Observatoire de Paris/LESIA (Paris), and Observatoire de Lyon, also supported by a grant from Labex OSUG@2020 (Investissements d'avenir - ANR10 LABX56). SPHERE is an instrument designed and built by a consortium consisting of IPAG (Grenoble, France), MPIA (Heidelberg, Germany), LAM (Marseille, France), LESIA (Paris, France), Laboratoire Lagrange (Nice, France), INAF - Osservatorio di Padova (Italy), Observatoire de Genève (Switzerland), ETH Zurich (Switzerland), NOVA (Netherlands), ONERA (France), and ASTRON (Netherlands), in collaboration with ESO. SPHERE was funded by ESO, with additional contributions from CNRS (France), MPIA (Germany), INAF (Italy), FINES (Switzerland), and NOVA (Netherlands). SPHERE received funding from the European Commission Sixth and Seventh Framework Programmes as part of the Optical Infrared Coordination Network for Astronomy (OPTICON) under grant number RII3-Ct-2004-001566 for FP6 (2004-2008), grant number 226604 for FP7 (2009-2012), and grant number 312430 for FP7 (2013-2016).

## Links

[1] https://sphere.osug.fr/spip.php?article45&lang=en
[2] http://cesam.lam.fr/spheretools

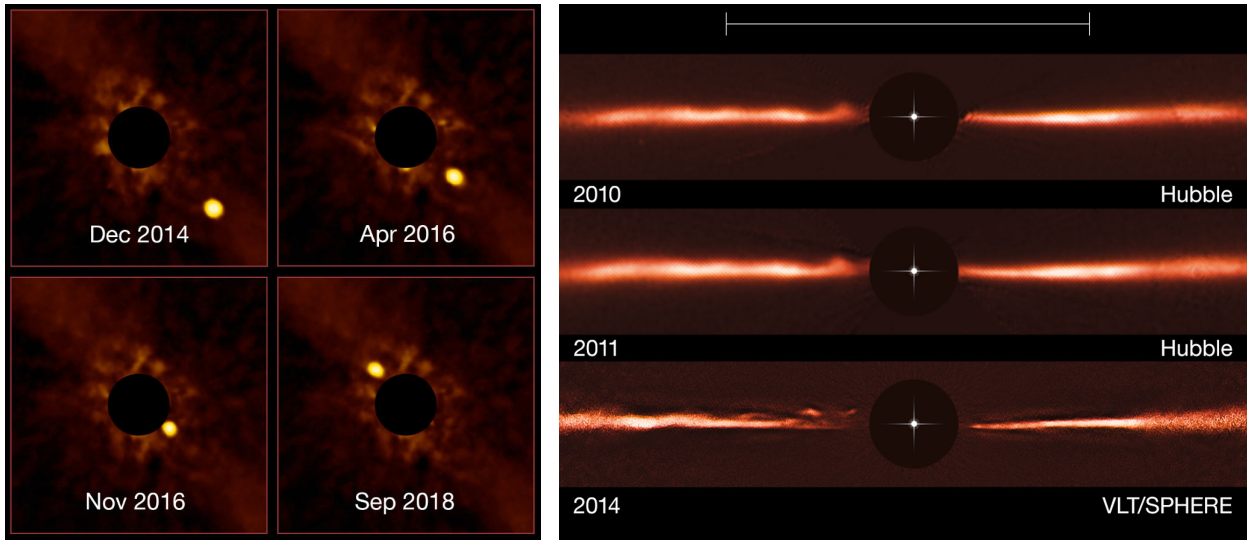

**Figure 1.**
SPHERE images at different epochs of the giant exoplanet β Pictoris b (left) and of arch-like disc features in the debris disc of AU Microscopii (right). High-precision relative astrometry is fundamental to measure the slow motions observed, from South-West to North-East for the planet and away from the star for the disc features. In the right panel, the scale bar at the top of the picture indicates the diameter of the orbit of the planet Neptune in the Solar System. Credits: ESO/Lagrange/SPHERE consortium and ESO/NASA/ESA.

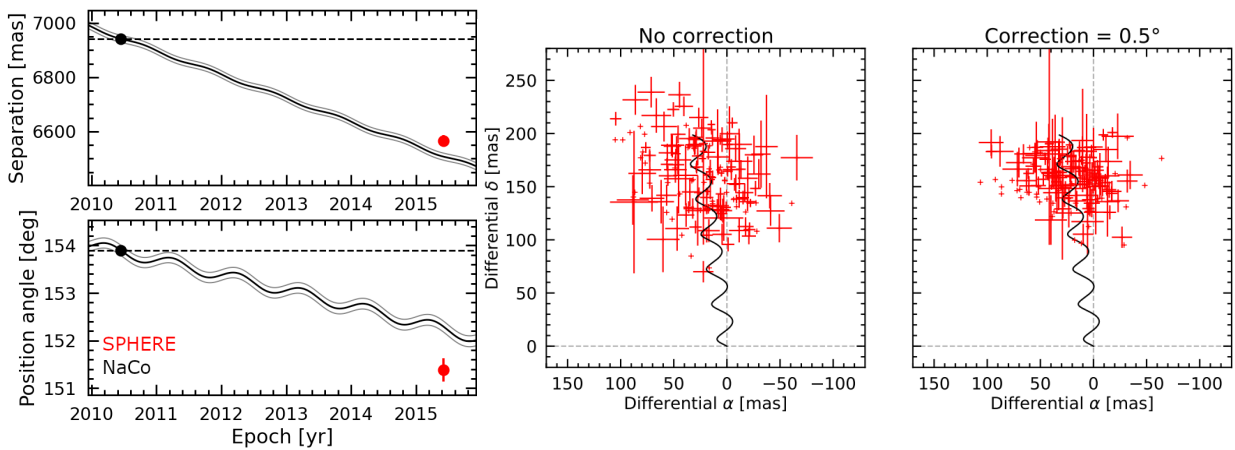

**Figure 2.**
Tests for companionship of companion candidates detected around stars with SPHERE and NaCo showing the importance of a good knowledge of the biases in relative astrometry for the interpretation. The left panel shows the temporal evolution of the separation to the star (top) and position angle relative to the North (bottom, counted positive from North to East) for a single candidate companion compared to the evolution for a stationary background contaminant (black curve with grey areas). The motion of the point source does not follow the stationary background track, suggesting a physical companion. However, underestimated systematic uncertainties could account for the discrepancies. The right panel shows the differential declination as a function of the differential right ascension for a star with many candidate companions (red data points) for two values of the assumed correction angle to the North. The black curves show the motion for a stationary background contaminant. If most candidate companions are assumed to be stationary background contaminants, a correction angle to the North is needed to make their motion compatible with the expected behaviour.

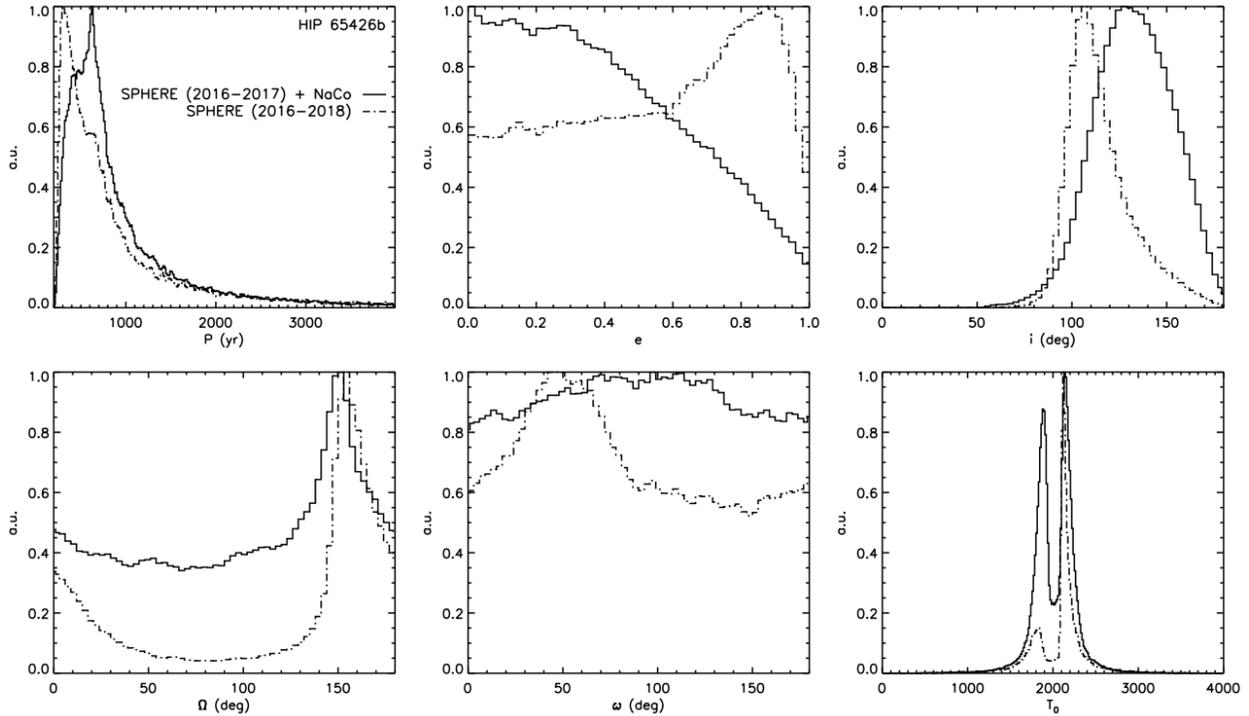

**Figure 3.**
Distributions of the orbital parameters of the exoplanet HIP65426b using two different sets of relative astrometric measurements. A good knowledge of the biases is mandatory to derive unbiased constraints. The panels show the period, eccentricity, inclination, longitude of the node, and argument and time at periapsis, from left to right and top to bottom.

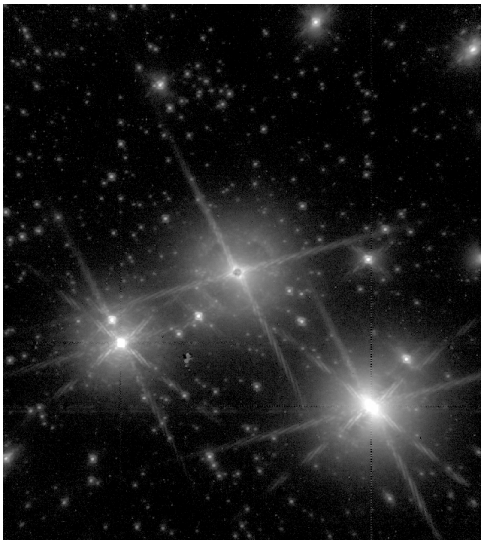

**Figure 4.**
SPHERE image of the 47 Tucanae field used for the astrometric calibration. The field of view is ~11'' on one side.

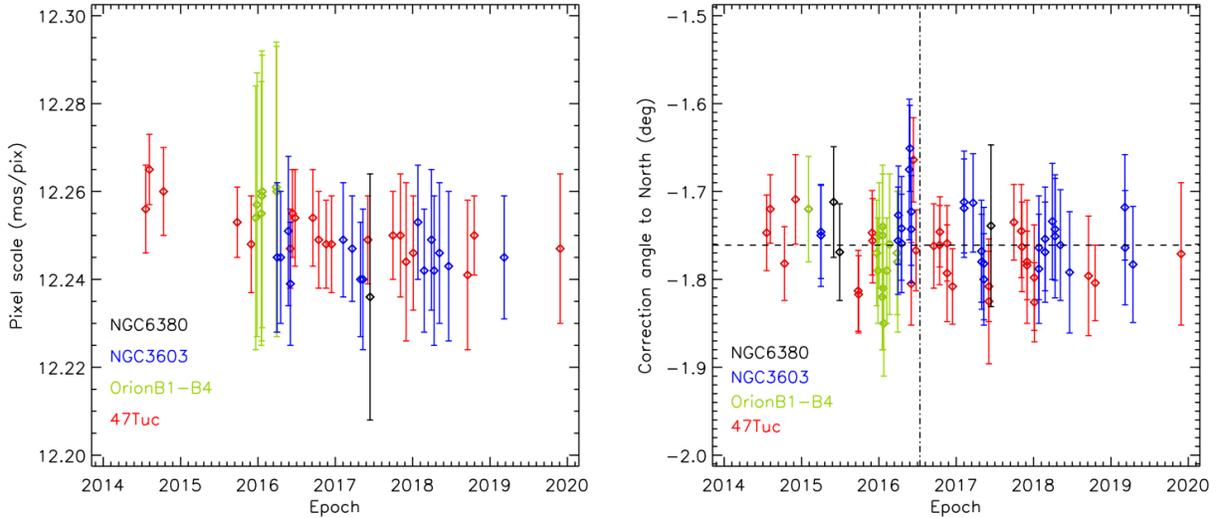

**Figure 5.**
Evolution of the pixel scale (left) and correction angle to the North (right) of SPHERE. A good instrument stability is mandatory for high-precision relative astrometry over time because it reduces potential systematic uncertainties. Fewer measurements are shown for the pixel scale because it depends on the filter and coronagraph configuration. For the right panel, the dotted-dashed vertical line indicates the epoch when the time reference issue was solved (Maire et al. 2016). All previous measurements were corrected. The dashed horizontal line shows the weighted mean of all the measurements.